\documentclass[aps,twocolumn]{revtex4}

\usepackage{graphics}
\usepackage{epsfig}

\usepackage{amsmath}
\usepackage{amssymb}
\usepackage{stmaryrd}

\def\ov#1{\overline{#1}}

\def\vb#1{\mbox{\boldmath$#1$}}
\def\pd#1#2{\frac{\partial #1}{\partial #2}}

\def\wh#1{\widehat{#1}}
\def\bdot{\,\vb{\cdot}\,}
\def\btimes{\,\vb{\times}\,}

\def\bhat{\wh{{\sf b}}}
\def\exd{{\sf d}}

\newcommand{\bc}{\begin{center}}
\newcommand{\ec}{\end{center}}
\newcommand{\bt}{\begin{tabbing}}
\newcommand{\et}{\end{tabbing}} 
\newcommand{\be}{\begin{eqnarray*}}
\newcommand{\ee}{\end{eqnarray*}}

\begin{document}

\title{On the validity of the guiding-center approximation in a magnetic dipole field}

\author{Alain J.~Brizard$^{1,a}$ and Danielle G.~Markowski$^{1,2}$}
\affiliation{$^{1}$Department of Physics, Saint Michael's College, Colchester, VT 05439, USA \\ $^{2}$Department of Physics and Astronomy, Clemson University, Clemson, SC 29634-0978, USA \\ $^{a}$Author to whom correspondence should be addressed: abrizard@smcvt.edu}

\begin{abstract}
The problem of the charged-particle motion in an axisymmetric magnetic-dipole geometry is used to assess the validity of Hamiltonian guiding-center theory, which includes higher-order corrections associated with guiding-center polarization induced by magnetic-field nonuniformity. When a magnetically-confined charged-particle orbit is regular (i.e., its guiding-center magnetic moment is adiabatically invariant), the guiding-center approximation, which conserves both energy and azimuthal canonical angular momentum, is shown to be faithful to the particle orbit when guiding-center polarization effects are taken into account.
\end{abstract}

\date{\today}

\maketitle

\section{Introduction}

Hamiltonian guiding-center theory  \cite{Littlejohn_1983,Cary_Brizard_2009} yields a reduced dynamical description in which the fast gyromotion of a charged particle about a local magnetic field line is asymptotically decoupled from the slow bounce and drift motions along and across magnetic-field lines. In particular, higher-order guiding-center corrections \cite{Burby_2013,Tronko_Brizard_2015} have been shown to play an important role in the faithful representation of regular charged-particle orbits in nonuniform magnetic fields. For example, the validity of the guiding-center approximation was recently explored in a straight magnetic field with constant gradient \cite{Brizard_2017}. There, it was shown that the guiding-center approximation remains valid even in the presence of strong gradients. See Ref.~\cite{Kabin_2021} for a recent discussion of the adiabatic invariance of the magnetic moment in a straight magnetic field with constant gradient.

Here, we investigate how a truncated guiding-center expression for an exact particle dynamical invariant can accurately represent this particle invariant. In previous work, the higher-order guiding-center corrections have been shown to be relevant for magnetically-confined energetic charged-particle orbits in realistic axisymmetric magnetic-field geometries \cite{Belova_2003}, where the faithful guiding-center representation of the toroidal angular momentum for these energetic particles required higher-order corrections (up to first order in magnetic-field nonuniformity).

The purpose of our present work is to explore how higher-order Hamiltonian guiding-center theory is faithful to charged single-particle dynamics in an axisymmetric magnetic dipole field, in which the azimuthal canonical angular momentum is an exact invariant. In particular, we compare the faithfulness of the truncated expressions of the guiding-center azimuthal angular momentum derived in the original work of Littlejohn \cite{Littlejohn_1983} and the recent work of Tronko and Brizard \cite{Tronko_Brizard_2015}, which includes effects due to guiding-center polarization \cite{Kaufman_1986,Brizard_2013}. 

The remainder of the paper is organized as follows. In Sec.~\ref{sec:dipole}, we present a brief description of the generic axisymmetric magnetic-dipole geometry considered in this work. While the magnetic dipole field considered here is a generic dipole field, we use the example of the Earth's geomagnetic dipole field \cite{Parks_2003} whenever we need explicit field parameters. In Sec.~\ref{sec:Newton}, we present the numerical solutions of the equations of motion for a charged particle (with mass $m$ and charge $q$) moving in an axisymmetric magnetic dipole field. These solutions satisfy the conservation laws of energy and azimuthal angular canonical momentum exactly. In addition, when the magnetic moment $\mu$ is a valid adiabatic invariant, the particle orbits are labeled by the invariants $({\cal E}, P_{\varphi},\mu)$ and they exhibit the expected three orbital time scales involving the fast gyromotion, the intermediate bounce motion, and the slow azimuthal drift-precession motion. In Sec.~\ref{sec:gc}, we present the guiding-center equations of motion expressed as Euler-Lagrange equations derived from a Lagrangian that contains higher-order guiding-center corrections. These equations conserve guiding-center energy and guiding-center azimuthal angular canonical momentum exactly. 

In Sec.~\ref{sec:validity}, the validity of the guiding-center approximation is explored in axisymmetric magnetic dipole geometry. In particular, we compare the standard guiding-center correction $\mu\,({\bf R} + \frac{1}{2}\,\tau\,\bhat)$, first derived by Littlejohn \cite{Littlejohn_1983}, where ${\bf R}$ denotes the gyrogauge vector field and the magnetic twist  $\tau \equiv \bhat\bdot\nabla\btimes\bhat$ is defined in terms of the magnetic-field unit vector field $\bhat$, with the guiding-center correction $\mu\,({\bf R} + \frac{1}{2}\,\nabla\btimes\bhat)$, which was derived more recently \cite{Tronko_Brizard_2015} by requiring that the guiding-center transformation yield the correct expression for the guiding-center polarization \cite{Kaufman_1986}. It will be shown that, in magnetic-dipole geometry (where the magnetic twist vanishes, $\tau \equiv 0$), the guiding-center correction $\mu\,({\bf R} + \frac{1}{2}\,\nabla\btimes\bhat)$ yields a guiding-center description that is faithful to the exact particle dynamics.

\section{\label{sec:dipole}Magnetic Dipole Geometry}

We begin by considering the problem of charged-particle motion in axisymmetric magnetic dipole geometry, where the magnetic field is represented by the generic expression
\begin{equation}
{\bf B} \;=\; \frac{B_{\rm e}\,r_{\rm e}^{3}}{r^{3}} \left( 2\,\sin\lambda\;\wh{r} \;-\; \cos\lambda\,\wh{\lambda} \right).
\label{eq:B_dip}
\end{equation}
Here, $B_{\rm e}$ denotes the strength of the equatorial magnetic field  at $(r,\lambda) = (r_{\rm e},0)$, and modified spherical coordinates $(r,\varphi,\lambda)$ are used (with the latitude angle $-\pi/2 \leq \lambda \leq \pi/2$ used instead of the polar angle $\theta = \pi/2 - \lambda$), with the unit vectors $(\wh{r}, \wh{\lambda} = \partial\wh{r}/\partial\lambda \equiv \wh{r}\btimes\wh{\varphi})$ on the poloidal plane:
\begin{equation}
\begin{array}{rcl}
\wh{r} & = & \cos\lambda \left( \cos\varphi\;\wh{\sf x} + \sin\varphi\;\wh{\sf y}\right) \;+\; \sin\lambda\;\wh{\sf z}, \\
\wh{\lambda} & = & -\;\sin\lambda \left( \cos\varphi\;\wh{\sf x} + \sin\varphi\;\wh{\sf y}\right) \;+\; \cos\lambda\;\wh{\sf z}.
\end{array}
\end{equation} 
The magnitude of the magnetic field \eqref{eq:B_dip} is expressed as
\begin{equation}
B(r,\lambda) \;=\; \frac{B_{\rm e}\,r_{\rm e}^{3}}{r^{3}}\;\sqrt{1 + 3\,\sin^{2}\lambda} \;\equiv\; B_{\rm e}\,\ov{B},
\label{eq:mag_dip}
\end{equation}
while the unit vector along the magnetic field is
\begin{equation}
\bhat(\lambda,\varphi) \;=\; \frac{2\,\sin\lambda\;\wh{r} \;-\; \cos\lambda\,\wh{\lambda}}{\sqrt{1 + 3\,\sin^{2}\lambda}}.
\label{eq:unit_dip}
\end{equation}
In the present work, dimensionless variables and fields are normalized to their equatorial values and are represented with an overbar (e.g., $\ov{r} = r/r_{\rm e}$). In addition, we will occasionally use the parameters associated with the geomagnetic dipole field \cite{Parks_2003}, where $B_{\rm e} = B_{\rm E}\,(r_{\rm E}/r_{\rm e})^{3}$, with $B_{\rm E} \simeq 3.12 \times 10^{-5}$ Tesla and $r_{\rm E} = 6378$ km, whenever explicit numerical values are needed.

The magnetic dipole field \eqref{eq:B_dip} may also be written in terms of the magnetic vector potential ${\bf A} = \psi\,\nabla\varphi$ as
\begin{equation}
{\bf B} \;=\; \nabla\btimes{\bf A} \;=\; \nabla\psi\btimes\nabla\varphi,
\label{eq:psi_dip}
\end{equation}
which guarantees that the magnetic dipole field is divergenceless: $\nabla\bdot{\bf B} = 0$, where the dipole magnetic flux is 
\begin{equation}
\psi(r,\lambda) \;=\; \psi_{\rm e}\,(r_{\rm e}/r)\;\cos^{2}\lambda \;\equiv\; \psi_{\rm e}\,\ov{\psi},
\label{eq:psi_def}
\end{equation}
with $\psi_{\rm e} = B_{\rm e}\,r_{\rm e}^{2}$.

Since the magnetic dipole field satisfies the condition ${\bf B}\bdot\nabla\psi = 0$, the magnetic field lines \eqref{eq:B_dip} lie on a constant-$\psi$ surface: $\psi = \psi_{\rm e}$ (i.e., $r = r_{\rm e}\,\cos^{2}\lambda$). On this constant-$\psi$ surface, the magnitude \eqref{eq:mag_dip} of the magnetic dipole field is a function of latitude $\lambda$ alone: $B|_{\psi}(\lambda) = B_{\rm e}\,(1 + 3\,\sin^{2}\lambda)^{\frac{1}{2}}/\cos^{6}\lambda$, which becomes infinite as $\lambda \rightarrow \pm\,\pi/2$.

Next, we note that, in the absence of current sources, the curl of the magnetic field \eqref{eq:B_dip} vanishes: $\nabla\btimes{\bf B} = 0$, which implies that
\begin{equation}
\bhat\btimes\nabla\ln B \;=\; \nabla\btimes\bhat \;=\; \bhat\btimes(\bhat\bdot\nabla\bhat) \;=\; K(\lambda)\;\nabla\varphi,
\label{eq:curl_b}
\end{equation}
where the magnetic curvature $\bhat\bdot\nabla\bhat = K\,\nabla\varphi\btimes\bhat$ is expressed in terms of the dimensionless latitude-dependent magnetic-curvature function
\begin{equation}
K(\lambda) \;=\; \frac{3\;(1 - \sin^{4}\lambda)}{(1 + 3\,\sin^{2}\lambda)^{\frac{3}{2}}}.
\label{eq:K_def}
\end{equation}
This magnetic-curvature function, which is a monotonic function of latitude $\lambda$: $0 \leq K \leq 3$, will play a key role in our discussion of the faithfulness of the guiding-center approximation.

\section{\label{sec:Newton}Phase-space Lagrangian Dynamics}

In order to derive equations of motion, we begin with the Lagrangian
\begin{equation}
L \;=\; \frac{q}{c}\,\psi\,\dot{\varphi} \;+\; \frac{m}{2} \left( \dot{r}^{2} + r^{2}\,\dot{\lambda}^{2} + r^{2}\,\cos^{2}\lambda\;\dot{\varphi}^{2}\right),
\label{eq:Big_gamma_def}
\end{equation}
where a dot denotes a time derivative and we used ${\bf A}\bdot\dot{\bf x} = \psi\,\dot{\varphi}$. From this Lagrangian, the Euler-Lagrange equations \cite{Brizard_2015} are expressed in terms of the spatial coordinates $(r,\lambda,\varphi)$ in dimensionless form:
\begin{eqnarray}
\ov{r}^{\prime\prime} &=&  \ov{r}\,\lambda^{\prime 2} \;+\; \varphi^{\prime}\cos^{2}\lambda \left(\ov{r}\,\varphi^{\prime} - \epsilon^{-1}/\ov{r}^{2}\right), \label{eq:rho_dot} \\
\lambda^{\prime\prime} &=& -\;\varphi^{\prime}\cos\lambda\sin\lambda \left(\varphi^{\prime} + \epsilon^{-1}/\ov{r}^{3}\right) \;-\; 2\,\ov{r}^{\prime}\lambda^{\prime}/\ov{r}, \label{eq:lambda_dot} \\
0 &=& \left( \ov{\psi} \;+\frac{}{} \epsilon\,\varphi^{\prime}\;\ov{r}^{2}\,\cos^{2}\lambda \right)^{\prime}, 
\label{eq:phi_dot}
\end{eqnarray}
where all time derivatives (denoted with a prime)
\begin{equation}
\frac{d}{d\ov{t}} \;\equiv\; \frac{1}{\epsilon\,\Omega_{\rm e}}\;\frac{d}{dt} 
\label{eq:epsilon}
\end{equation}
 are now normalized to the bounce-motion time scale associated with the periodic motion along the latitude $\lambda$, i.e., $\dot{\varphi} = \epsilon\,\Omega_{\rm e}\,\varphi^{\prime}$, where $\epsilon\,\Omega_{\rm e} \ll \Omega_{\rm e} = qB_{\rm e}/mc$. For example, using the geomagnetic dipole field at $r_{\rm e} = 2\,r_{\rm E}$, the fast gyromotion time scale is $2\pi/\Omega_{\rm e} \simeq 9\,\mu{\rm sec}$ for electrons and  $2\pi/\Omega_{\rm e} \simeq 17$ msec for protons. In addition, at $r_{\rm e} = 2\,r_{\rm E}$, the particle velocity is normalized to $\epsilon\,r_{\rm e}\Omega_{\rm e} \simeq 16\,\epsilon\,c$ (for protons), which implies that $\epsilon \ll 1/16$ in order to satisfy the non-relativistic assumption. In this work, we shall use $\epsilon = 1/50$, which places the bounce period for protons on the scale of a few seconds.

We note that, according to Eq.~\eqref{eq:phi_dot}, the covariant azimuthal component of the canonical momentum ${\bf P} = (q/c)\,\psi\,\nabla\varphi + m\,\dot{\bf x}$:
\begin{eqnarray}
P_{\varphi} &=& {\bf P}\bdot\partial{\bf x}/\partial\varphi \;=\; (q/c)\,\psi \;+\; m\,\dot{\varphi}\,|\partial{\bf x}/\partial\varphi|^{2} \label{eq:P_phi} \\
  &=& (q/c)\psi_{\rm e}\left( \ov{r}^{-3} + \epsilon\,\varphi^{\prime}\right) \ov{r}^{2}\,\cos^{2}\lambda \;\equiv\; (q/c)\psi_{\rm e}\,\ov{P}_{\varphi}
\nonumber
\end{eqnarray}
is an exact dynamical invariant, which follows from the azimuthal symmetry of the magnetic dipole field.

\begin{widetext}
\bc
 \begin{figure}
\epsfysize=1.3in
\epsfbox{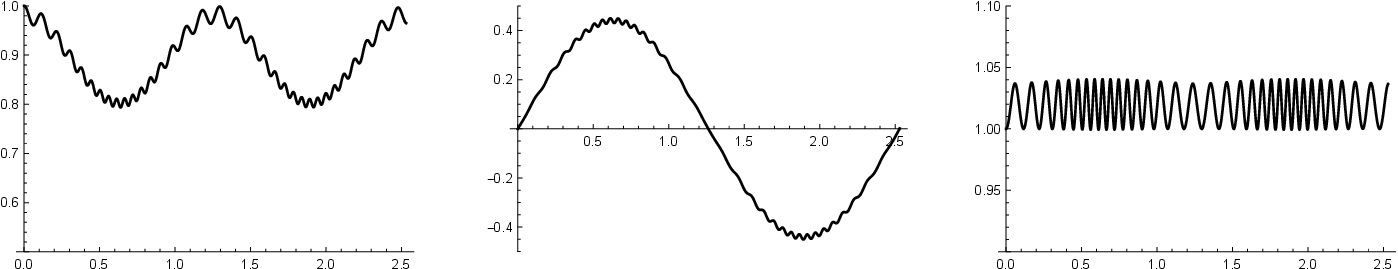}
\caption{Plots of (left) the normalized radial position $r(\epsilon\,\Omega_{\rm e} t)/r_{\rm e} = \ov{r}(\ov{t})$, (center) the latitude $\lambda(\ov{t})$, and (right) the normalized magnetic flux $\psi(\epsilon\,\Omega_{\rm e}t)/\psi_{\rm e} = \ov{\psi}(\ov{t}) = \cos^{2}\lambda(\ov{t})/\ov{r}(\ov{t})$ over one (normalized) bounce period $0 \leq  \ov{t} = \epsilon\,\Omega_{\rm e}t \leq T_{\rm b} \simeq 2.5$ for $\epsilon = 1/50$, using the equations of motion \eqref{eq:rho_dot}-\eqref{eq:phi_dot} with the initial conditions \eqref{eq:initial}.}
\label{fig:rho_lambda_psi}
\end{figure}
\ec
\end{widetext}

\begin{figure}
\epsfysize=1.8in
\epsfbox{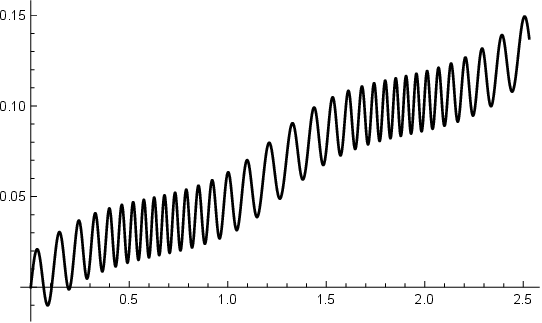}
\caption{Plot of the azimuthal angle $\varphi(\ov{t})$ over one bounce period $0 \leq  \ov{t} = \epsilon\,\Omega_{\rm e}t \leq T_{\rm b} \simeq 2.5$ for $\epsilon = 1/50$, using the equations of motion \eqref{eq:rho_dot}-\eqref{eq:phi_dot} with the initial conditions \eqref{eq:initial}.}
\label{fig:phi}
\end{figure}

\begin{figure}
\epsfysize=1.8in
\epsfbox{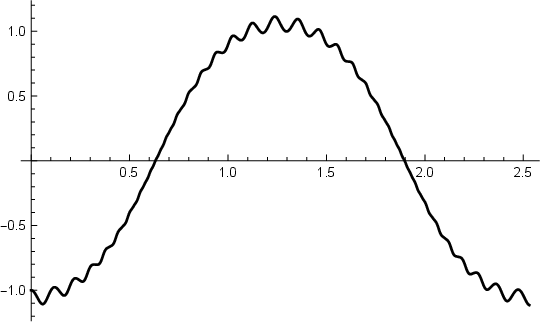}
\caption{Plot of the normalized parallel particle velocity $v_{\|}(\epsilon\,\Omega_{\rm e}t)/(\epsilon\,r_{\rm e}\Omega_{\rm e}) = \ov{v}_{\|}(\ov{t})$, with $v_{\|}$ given by Eq.~\eqref{eq:v_par||}, over a bounce period $0 \leq  \ov{t} = \epsilon\,\Omega_{\rm e}t \leq T_{\rm b} \simeq 2.5$ for $\epsilon = 1/50$ and the initial conditions \eqref{eq:initial}.}
\label{fig:v_par}
\end{figure}

\begin{figure}
\epsfysize=3.2in
\epsfbox{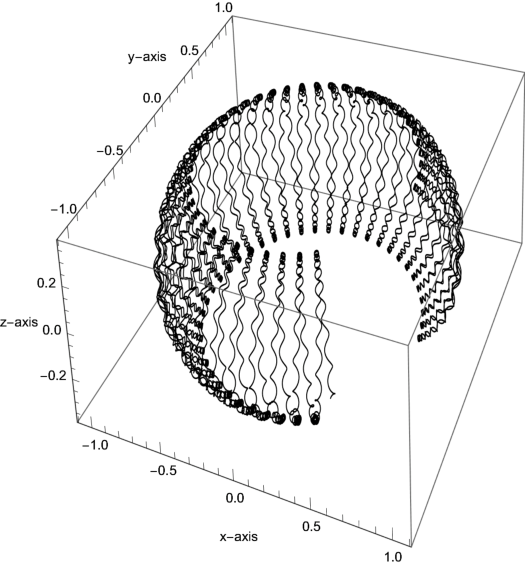}
\caption{Normalized particle orbit (over several bounce periods) showing the fast gyromotion, the intermediate bounce motion (predominantly along the $z$-axis), and the slow azimuthal precession motion (projected onto the $(x,y)$ equatorial plane).}
\label{fig:orbit_par}
\end{figure}

Figures \ref{fig:rho_lambda_psi}-\ref{fig:orbit_par} show the numerical solutions of the equations of motion \eqref{eq:rho_dot}-\eqref{eq:phi_dot} over one normalized bounce period $T_{\rm b} \simeq 2.5$, with initial conditions 
\begin{eqnarray}
(\ov{r}_{0}, \lambda_{0}, \varphi_{0}) &=& (1,0,0), \nonumber \\
 && \label{eq:initial} \\
 (\ov{r}_{0}^{\prime}, \lambda_{0}^{\prime}, \varphi_{0}^{\prime}) &=& (0,1,1), 
 \nonumber 
 \end{eqnarray}
 so that a particle orbit begins on the equatorial plane. These figures show the fast gyromotion oscillations and the slow bounce-motion modulation, as a particle oscillates between latitude values $-\lambda_{\rm b} \leq \lambda \leq \lambda_{\rm b}$, as can be seen in the center frame of Fig.~\ref{fig:rho_lambda_psi}. We note that because of finite-gyroradius effects, a charged particle does not move on a constant-$\psi$ surface (right frame of  Fig.~\ref{fig:rho_lambda_psi}) and its departure is described by the gyroradius vector
\begin{equation}
\vb{\rho}_{0} = \frac{\bhat}{\Omega}\btimes{\bf v} = \frac{1}{B\Omega}\left( \dot{\psi}\;\nabla\varphi - \dot{\varphi}\;\nabla\psi\right) \equiv \epsilon\,r_{\rm e}\,\ov{\vb{\rho}}_{0},
\label{eq:rho}
\end{equation}
where $|\ov{\vb{\rho}}_{0}| \;=\; [2\mu B_{\rm e}/(\epsilon^{2}m r_{\rm e}^{2}\Omega_{\rm e}^{2})]^{\frac{1}{2}} = \ov{\mu}^{\frac{1}{2}}$. In Eq.~\eqref{eq:rho}, the ordering parameter $\epsilon$, defined in the time-scale ordering \eqref{eq:epsilon}, appears as $|\vb{\rho}_{0}|/r_{\rm e} = \epsilon\,\ov{\mu}^{\frac{1}{2}} \ll 1.$

The kinetic energy ${\cal E} = \frac{1}{2}\,m|{\bf v}|^{2} \equiv \frac{1}{2}\epsilon^{2}mr_{\rm e}^{2}\Omega_{\rm e}^{2}\,\ov{\cal E}$ is also an exact dynamical invariant since $\dot{\cal E} = m\dot{\bf v}\bdot{\bf v} = (q/c)\,({\bf v}\btimes{\bf B})\bdot{\bf v} = 0$. Here, it is possible to separate the parallel and perpendicular components of the particle velocity as follows. First, we define the parallel particle velocity 
\begin{equation}
v_{\|} = {\bf v}\bdot\bhat \;=\; \frac{2\,\dot{r}\,\sin\lambda \;-\; r\,\dot{\lambda}\,\cos\lambda}{\sqrt{1 + 3\,\sin^{2}\lambda}} \equiv \epsilon\,r_{\rm e}\Omega_{\rm e}\,\ov{v}_{\|}
\label{eq:v_par||} 
\end{equation}
so that, according to Eq.~\eqref{eq:initial}, the initial normalized parallel velocity is $\ov{v}_{\|0} = -\lambda^{\prime}_{0} = -1$. The perpendicular velocity can be expressed as
\begin{eqnarray}
{\bf v}_{\bot} &=& (\bhat\btimes{\bf v})\btimes\bhat \;=\; \frac{\dot{\psi}}{|\nabla\psi|}\;\wh{\psi} \;+\;  \frac{\dot{\varphi}}{|\nabla\varphi|}\;\wh{\varphi} \nonumber \\
 &\equiv& \epsilon\,r_{\rm e}\Omega_{\rm e}\,\ov{\vb{\rho}}_{0}\btimes\bhat,
\label{eq:v_par_perp}
\end{eqnarray}
with the initial normalized condition $\ov{\bf v}_{\bot 0} = \varphi_{0}^{\prime}\,\wh{\varphi}_{0} = \wh{\varphi}_{0}$. According to the initial conditions \eqref{eq:initial}, our numerical particle orbit begins on the equatorial plane with a pitch angle $\alpha_{0} = 
\arctan(|{\bf v}_{\bot0}|/v_{\|0}) = 3\pi/4$ with normalized total energy $\ov{\cal E} = 2$.

The normalized parallel velocity is shown in Fig.~\ref{fig:v_par} over one bounce period, where large-amplitude gyro-oscillations are observed as the particle moves near the equatorial plane. The particle reaches a bounce turning-point $\pm\lambda_{\rm b}$ as $v_{\|}$ vanishes and reverses sign. The 3D particle orbit over several bounce periods is shown in Fig.~\ref{fig:orbit_par}, where we observe the fast gyromotion (spiraling orbit about a single magnetic-field line), intermediate bounce motion (projected on the $z$-axis) between turning points where the parallel velocity \eqref{eq:v_par||} vanishes.

\section{\label{sec:gc}Guiding-center Phase-space Lagrangian Dynamics}

In order to remove the fast gyromotion time scale from the equations of motion \eqref{eq:rho_dot}-\eqref{eq:phi_dot}, as seen in Figs.~\ref{fig:rho_lambda_psi}-\ref{fig:orbit_par}, and extract slower orbital bounce and azimuthal drift-precession time scales, we proceed with the guiding-center representation for charged-particle motion in a nonuniform magnetic field (see Refs.~\cite{Littlejohn_1983,Cary_Brizard_2009,Tronko_Brizard_2015}). 

Using the guiding-center coordinates $({\bf X}, P_{\|},\mu, \zeta)$, the phase-space Lagrangian \eqref{eq:Big_gamma_def} is transformed into the guiding-center phase-space Lagrangian \cite{Tronko_Brizard_2015}
\begin{equation}
L_{\rm gc} \;=\; \left( \frac{q}{c}\,{\bf A} + P_{\|}\;\bhat \right)\bdot\dot{{\bf X}} + J\,\left( \dot{\zeta} - {\bf R}^{*}\bdot\dot{{\bf X}} \right) - H_{\rm gc},
\label{eq:ovgamma_gc}
\end{equation}
where ${\bf X} = (R,\Phi,\Lambda)$ denotes the guiding-center position, $P_{\|}$ denotes the parallel guiding-center momentum, and the guiding-center gyroaction $J = \mu\,B_{\rm e}/\Omega_{\rm e} = \mu\,(mc/q)$ is canonically conjugate to the guiding-center gyroangle $\zeta$, and the guiding-center Hamiltonian is $H_{\rm gc} = P_{\|}^{2}/2m + \mu\,B$. The guiding-center position ${\bf X}({\bf x},{\bf v}) = {\bf x} - \vb{\rho}_{0} - \cdots$ is defined with the help of the (lowest-order) gyroradius vector \eqref{eq:rho}, so that the guiding-center spherical coordinates are 
\begin{equation}
\left. \begin{array}{rcl}
R &=& r \;-\; \epsilon\,\ov{\vb{\rho}}_{0}\bdot\ov{\nabla} r \;+\; \cdots \\
\Lambda &=& \lambda \;-\; \epsilon\,\ov{\vb{\rho}}_{0}\bdot\ov{\nabla}\lambda \;+\; \cdots \\
\Phi &=& \varphi \;-\; \epsilon\,\ov{\vb{\rho}}_{0}\bdot\ov{\nabla}\varphi \;+\; \cdots
\end{array} \right\},
\label{eq:X_gc}
\end{equation}
where we used the ordering $\vb{\rho}_{0}\bdot\nabla = \epsilon\,\ov{\vb{\rho}}_{0}\bdot\ov{\nabla}$ and higher-order corrections depend on the magnetic-field nonuniformity. The guiding-center transformation from the local momentum coordinates $(p_{\|} = \bhat\bdot{\bf p}, J_{0} = \mu_{0}\,B_{\rm e}/\Omega_{\rm e}, \zeta_{0})$ to the guiding-center momentum coordinates $(P_{\|},J \equiv \mu\,B_{\rm e}/\Omega_{\rm e},\zeta)$, on the other hand, depends on magnetic-field nonuniformity \cite{Cary_Brizard_2009}. In the present work, the first-order correction $\mu_{1}$, which is derived in the Appendix, will be needed in order to verify the validity of the guiding-center approximation.

\begin{figure}
\epsfysize=2in
\epsfbox{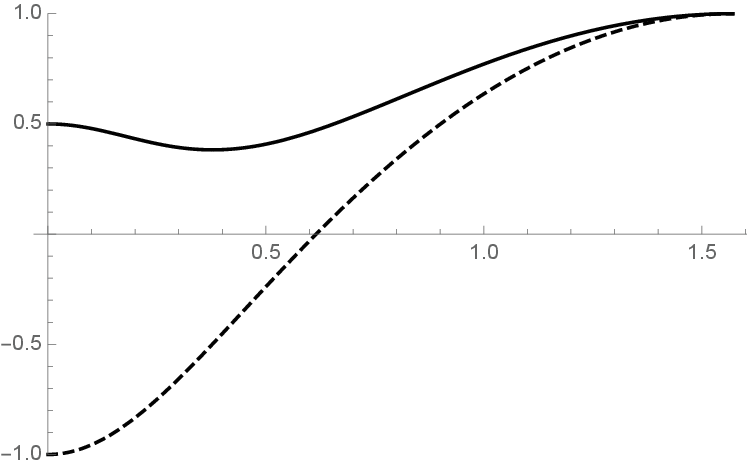}
\caption{Plots of $b_{z}^{*}$ (solid) and $b_{z}$ (dashed) as functions of the guiding-center latitude angle $0 \leq \Lambda \leq \pi/2$. The impact of the dimensionless magnetic curvature $b_{z}^{*} - b_{z} = \frac{1}{2}\,K$ is clearly seen with $b_{z}^{*} > 0$ in the range $0 \leq \Lambda \leq \pi/2$.}
\label{fig:b_z}
\end{figure}

The vector functions ${\bf A} = \Psi\,\nabla\Phi$ and ${\bf B} = B\,\bhat$ in Eq.~\eqref{eq:ovgamma_gc} are evaluated at the guiding-center position ${\bf X}$. We note that, as a result of the guiding-center transformation, the magnetic moment $\mu$ is an invariant of the guiding-center dynamics, while it is not an invariant of the particle dynamics (the breakdown of the adiabatic invariance of the magnetic moment is briefly discussed in App.~\ref{sec:gc_higher}). In Eq.~\eqref{eq:ovgamma_gc}, the guiding-center correction is expressed in terms of the guiding-center vector field
\begin{equation}
{\bf R}^{*} \;=\; {\bf R} \;+\; \left\{ \begin{array}{lr}
\frac{1}{2}\,\nabla\btimes\bhat \;\equiv\; \frac{1}{2}\,K(\Lambda)\,\nabla\Phi & ({\sf A}) \\
 & \\
 \frac{1}{2}\,(\bhat\bdot\nabla\btimes\bhat)\;\bhat \;\equiv\; 0 & ({\sf B})
 \end{array} \right.
 \label{eq:Wstar_def}
 \end{equation} 
defined in terms of the gyrogauge vector field \cite{Littlejohn_1983} ${\bf R} = \nabla\wh{1}\bdot\wh{2}$, which is expressed in terms of the orthogonal unit vector fields $(\wh{1}, \wh{2}, \bhat = \wh{1}\btimes\wh{2})$, and a higher-order correction that either involves ({\sf A}) magnetic curvature \cite{Tronko_Brizard_2015} or ({\sf B}) a correction that vanishes in magnetic-dipole geometry \cite{Littlejohn_1983}. In magnetic-dipole geometry, where we can use $\wh{1} = \wh{\Phi}$ and $\wh{2} = \bhat\btimes\wh{\Phi}$, we find ${\bf R} = \nabla\Phi\,(\partial\wh{\Phi}/\partial\Phi)\bdot(\bhat\btimes\wh{\Phi}) = b_{z}(\Lambda)\,\nabla\Phi$, where $\partial\wh{\Phi}/\partial\Phi = \wh{\sf z}\btimes\wh{\Phi}$ and $b_{z} \equiv \bhat\bdot\wh{\sf z}$, while the dimensionless magnetic curvature $K$ is defined in Eq.~\eqref{eq:K_def}. Hence, the vector field ${\bf R}^{*} = b_{z}^{*}\;\nabla\Phi$ is expressed in terms of the curvature-modified function
\begin{eqnarray}
b_{z}^{*} &\equiv& b_{z} + \frac{1}{2}\,K \;=\; \frac{9\,\sin^{4}\Lambda - 1}{(1 + 3\,\sin^{2}\Lambda)^{\frac{3}{2}}} \;+\; \frac{3}{2} \frac{(1 - \sin^{4}\Lambda)}{(1 + 3\,\sin^{2}\Lambda)^{\frac{3}{2}}} \nonumber \\
 &=& \frac{(15\,\sin^{4}\Lambda + 1)}{2\,(1 + 3\,\sin^{2}\Lambda)^{\frac{3}{2}}}.
\label{eq:bz_*}
\end{eqnarray}
Figure \ref{fig:b_z} shows that, while $b_{z}(\Lambda)$ remains negative until $\Lambda > \arcsin(1/\sqrt{3}) \simeq \pi/5$, the curvature-modified function $b_{z}^{*}(\Lambda)$ remains positive in the range $0 \leq \Lambda \leq \pi/2$. Hence, the difference between $b_{z}^{*}$ and $b_{z}$ is seen to be small only for $\Lambda > \pi/3$.

\subsection{Guiding-center Euler-Lagrange equations}

The guiding-center Euler-Lagrange equations associated with variations of the guiding-center Lagrangian \eqref{eq:ovgamma_gc} in $({\bf X}, P_{\|},\mu, \zeta)$ yield, respectively,
\begin{eqnarray}
\dot{P}_{\|}\,\bhat -  \frac{q}{c}\,\dot{\bf X}\btimes{\bf B}^{*} - \dot{J}\,{\bf R}^{*} & = & -\,\nabla H_{\rm gc} = -\,\mu\nabla B, \label{eq:EL_X} \\
\bhat\bdot\dot{\bf X} & = & \partial H_{\rm gc}/\partial P_{\|} = P_{\|}/m, \label{eq:EL_q} \\
\dot{\zeta} \;-\; \dot{\bf X}\bdot{\bf R}^{*} & = & \partial H_{\rm gc}/\partial J = \Omega, \label{eq:EL_J} \\
\dot{J} & = & -\;\partial H_{\rm gc}/\partial\zeta = 0, \label{eq:EL_zeta}
\end{eqnarray}
where a dot denotes a standard time derivative and the curvature-modified magnetic field
\begin{eqnarray}
{\bf B}^{*} &=& \nabla\btimes\left[ {\bf A} \;+\; (c/q) \left(P_{\|}\;\bhat \;-\frac{}{} \epsilon\,J\,{\bf R}^{*}\right) \right] \nonumber \\
 &=& \nabla\Psi^{*}\btimes\nabla\Phi \;+\; (c/q)\;P_{\|}K(\Lambda)\,\nabla\Phi
\label{eq:B_*}
\end{eqnarray}
is defined in terms of the gyroaction-corrected magnetic flux
\begin{equation}
\Psi^{*} \;=\; \Psi(R,\Lambda) \;-\; (c/q)\,J\,b_{z}^{*}(\Lambda). 
\label{eq:Psi_*}
\end{equation}
Although Eq.~\eqref{eq:EL_zeta} implies that the guiding-center gyroaction $J$ is a guiding-center invariant (since the guiding-center Hamiltonian $H_{\rm gc}$ is independent of the gyroangle $\zeta$ up to second order in magnetic-field nonuniformity), it is not an exact particle invariant.

Making use of Eq.~\eqref{eq:EL_zeta}, Eqs.~\eqref{eq:EL_X}-\eqref{eq:EL_q} yield the reduced guiding-center equations of motion
\begin{eqnarray}
\dot{\bf X} & = & \frac{P_{\|}}{m}\;\frac{{\bf B}^{*}}{B_{\|}^{*}} \;+\; \frac{c\bhat}{qB_{\|}^{*}}\btimes\mu\,\nabla B \label{eq:X_dot} \\
 &=& \frac{c}{qB_{\|}^{*}}\left(\mu\,B + \frac{P_{\|}^{2}}{m}\right) K\,\nabla\Phi \;+\; \frac{P_{\|}}{mB_{\|}^{*}}\,\nabla\Psi^{*}\btimes
\nabla\Phi,  \nonumber \\
\dot{P}_{\|} & = & -\;\frac{{\bf B}^{*}}{B_{\|}^{*}}\bdot\mu\nabla B \;=\; -\;\frac{1}{B_{\|}^{*}}\;\nabla\Psi^{*}\btimes\nabla\Phi\bdot\mu\nabla B, \label{eq:q_dot}
\end{eqnarray}
where we used Eq.~\eqref{eq:curl_b} and the definition
\begin{eqnarray}
B_{\|}^{*} &\equiv& \bhat\bdot{\bf B}^{*} \;=\; \bhat\bdot\nabla\Psi^{*}\btimes\nabla\Phi  \label{eq:B_star} \\
 &=& \frac{-\,1}{{\cal J}\sqrt{1 + 3\,\sin^{2}\Lambda}}\left( 2\,\sin\Lambda\;\pd{\Psi^{*}}{\Lambda} + R\,\cos\Lambda\;\pd{\Psi^{*}}{R}\right),
\nonumber
\end{eqnarray}
where ${\cal J} = R^{2}\,\cos\Lambda$ is the Jacobian for the spatial coordinates $(R,\Phi,\Lambda)$. In Eq.~\eqref{eq:q_dot}, we used the fact that the magnitude of the magnetic dipole field $B$ is independent of the guiding-center azimuthal angle $\Phi$ (i.e., $\nabla\Phi\bdot\nabla B = 0$). We note that $P_{\|} = m\dot{\bf X}\bdot\bhat$ is indeed the guiding-center parallel momentum and the fast gyromotion, now described by Eq.~\eqref{eq:EL_J}, is decoupled from the slow guiding-center dynamics \eqref{eq:X_dot}-\eqref{eq:q_dot}. 

In addition, we note that, while the guiding-center energy ${\cal E} = P_{\|}^{2}/2m + \mu\,B$ is an exact guiding-center invariant:
\[ \dot{\cal E} \;=\; P_{\|}\,\dot{P}_{\|}/m \;+\; \mu\,\dot{\bf X}\bdot\nabla B \;=\; 0, \]
the guiding-center azimuthal canonical angular momentum 
\begin{equation}
P_{{\rm gc}\Phi} \;=\; (q/c)\,\Psi^{*} \;=\; (q/c)\,\Psi \;-\; J\;b_{z}^{*}
\label{eq:Pgc_phi}
\end{equation}
 is also an exact guiding-center invariant, since
\begin{equation}
\dot{\Psi}^{*} \;=\; \dot{\bf X}\bdot\nabla\Psi^{*} \;=\; \dot{R}\;\pd{\Psi^{*}}{R} \;+\; \dot{\Lambda}\;\pd{\Psi^{*}}{\Lambda} \;=\; 0,
\end{equation}
where, from Eq.~\eqref{eq:X_dot}, we used
\begin{eqnarray}
\dot{R} & = & \dot{\bf X}\bdot\nabla R \;=\; \frac{P_{\|}}{mB_{\|}^{*}}\,\nabla\Psi^{*}\btimes\nabla\Phi\bdot\nabla R \nonumber \\
 &=&  -\;\frac{P_{\|}}{m{\cal J}B_{\|}^{*}}\,\pd{\Psi^{*}}{\Lambda}, \label{eq:R_dot} \\
\dot{\Lambda} & = & \dot{\bf X}\bdot\nabla\Lambda \;=\; \frac{P_{\|}}{mB_{\|}^{*}}\,\nabla\Psi^{*}\btimes\nabla\Phi\bdot\nabla \Lambda \nonumber \\
 &=& \frac{P_{\|}}{m{\cal J}B_{\|}^{*}}\,\pd{\Psi^{*}}{R}. \label{eq:Lambda_dot}
\end{eqnarray}
Lastly, the guiding-center azimuthal angular velocity
\begin{equation}
\dot{\Phi} \;=\; \dot{\bf X}\bdot\nabla\Phi \;=\; \frac{c}{qB_{\|}^{*}}\left(\mu\,B + \frac{P_{\|}^{2}}{m}\right) K\,|\nabla\Phi|^{2} \label{eq:Phi_dot}
\end{equation}
is entirely driven by magnetic curvature, while the guiding-center parallel force equation \eqref{eq:q_dot} is driven by the magnetic-field gradient.

\subsection{Normalized guiding-center equations}

In order to extract explicit orbital time scales from the guiding-center equations of motion \eqref{eq:q_dot} and \eqref{eq:R_dot}-\eqref{eq:Phi_dot}, we now introduce the following normalizations. First, we introduce $B = B_{\rm e}\,\ov{B}$, 
$R = r_{\rm e}\,\rho$, and $\Psi = B_{\rm e}r_{\rm e}^{2}\,\ov{\Psi}$, so that
\begin{equation}
\Psi^{*} \;=\; B_{\rm e}r_{\rm e}^{2} \left( \ov{\Psi} \;-\; \frac{\epsilon^{2}}{2}\;\ov{\mu}\, b_{z}^{*}(\Lambda) \right) \;\equiv\; B_{\rm e}r_{\rm e}^{2}\,\ov{\Psi}^{*},
\label{eq:Psi_ov}
\end{equation}
where $\ov{B} = \rho^{-3}(1 + 3\sin^{2}\Lambda)^{\frac{1}{2}}$ and $\ov{\Psi} = \rho^{-1}\cos^{2}\Lambda$. Next, the normalized guiding-center motion in the poloidal $(\rho,\Lambda)$-plane is expressed as
\begin{equation}
\left( \dot{\rho},\; \dot{\Lambda} \right) \;=\; \frac{\nu_{\rm e}\,\ov{P}_{\|}}{\ov{\cal J}\ov{B}_{\|}^{*}}\,\left(-\;\pd{\ov{\Psi}^{*}}{\Lambda}, \pd{\ov{\Psi}^{*}}{\rho}\right), 
\label{eq:ov_dot}
\end{equation}
where we introduced the energy-dependent bounce frequency 
\begin{equation}
\nu_{\rm e} \;\equiv\; \sqrt{\frac{2\,\mu B_{\rm e}}{m r_{\rm e}^{2}}} \;\equiv\; \epsilon\,\Omega_{\rm e}\,\ov{\mu}^{\frac{1}{2}}.
\label{eq:nu_def}
\end{equation}
Here, we introduced the normalized guiding-center parallel momentum $\ov{P}_{\|} = P_{\|}/(mr_{\rm e}\nu_{\rm e})$, which satisfies the normalized guiding-center parallel-force equation
\begin{equation}
\dot{\ov{P}}_{\|} \;=\; -\;\frac{\nu_{\rm e}}{2\ov{\cal J}\ov{B}_{\|}^{*}} \left( \pd{\ov{\Psi}^{*}}{\rho}\,\pd{\ov{B}}{\Lambda} \;-\; \pd{\ov{\Psi}^{*}}{\Lambda}\,\pd{\ov{B}}{\rho} \right),
\label{eq:ovP_dot}
\end{equation}
so that the total guiding-center kinetic energy becomes
\begin{equation}
{\cal E} \;=\; \mu\,B_{\rm e} \left( \ov{B} \;+\; \ov{P}_{\|}^{2}\right) \;=\; \mu\,B_{\rm t} \;\equiv\; \mu B_{\rm e}\;\ov{B}_{\rm t},
\label{eq:E_gc}
\end{equation}
where $B_{\rm t}$ denotes the magnetic field at which the parallel guiding-center momentum vanishes. We easily see that $|\ov{P}_{\|}| = \sqrt{\ov{B}_{t} - \ov{B}}$ (which has a maximum at the equator $|\ov{P}_{\|{\rm e}}| = \sqrt{\ov{B}_{\rm t} - 1}$), where $1 \leq \ov{B} \leq \ov{B}_{\rm t} = {\cal E}/(\mu B_{\rm e})$, and the poloidal guiding-center motion \eqref{eq:ov_dot} also comes to a stop at a turning point. For this reason, the equations \eqref{eq:ov_dot} represent the periodic bounce motion between turning points located symmetrically in the northern and southern hemispheres. In an axisymmetric magnetic dipole field \cite{Parks_2003,Duthoit_2010}, the turning points $\pm\,\Lambda_{\rm b}$ are determined from the roots of the equation
\begin{equation}
\cos^{6}\Lambda_{\rm b} \;-\; \sin^{2}\alpha_{0}\,\sqrt{1 + 3\,\sin^{2}\Lambda_{\rm b}} \;=\; 0, 
\label{eq:lambda_b}
\end{equation}
where $\alpha_{0}$ denotes the initial pitch angle. The normalized guiding-center azimuthal motion is expressed as
\begin{equation}
\dot{\Phi} \;=\; \frac{\nu_{\rm e}^{2}}{2\,\Omega_{\rm e}\,\ov{B}_{\|}^{*}} \left(\ov{B} \;+\; 2\,\ov{P}_{\|}^{2}\right)\,K\,|\ov{\nabla}\Phi|^{2},
\label{eq:ovPhi_dot}
\end{equation}
which does not stop at a turning point since $\ov{B} + 2\,\ov{P}_{\|}^{2} = 2\,\ov{B}_{\rm t} - \ov{B} \geq \ov{B}_{\rm t}$ does not vanish.

We conclude this Section by noting that Eqs.~\eqref{eq:ov_dot} and \eqref{eq:ovPhi_dot} imply that time derivatives can be normalized to the bounce time scale $\ov{t} \equiv \nu_{\rm e}t$, so that we obtain the normalized guiding-center equations of motion
\begin{eqnarray}
\left( \rho^{\prime}(\ov{t}),\; \Lambda^{\prime}(\ov{t}) \right) & = & \frac{\ov{P}_{\|}}{\ov{\cal J}\ov{B}_{\|}^{*}}\,\left(-\;\pd{\ov{\Psi}^{*}}{\Lambda}, \pd{\ov{\Psi}^{*}}{\rho}\right), \label{eq:RLambda_prime} \\
\Phi^{\prime}(\ov{t}) & = & \frac{\nu_{\rm e}}{2\,\Omega_{\rm e}\,\ov{B}_{\|}^{*}} \left(\ov{B} \;+\; 2\,\ov{P}_{\|}^{2}\right)\,K\,|\ov{\nabla}\Phi|^{2}, \label{eq:Phi_prime} \\
\ov{P}_{\|}^{\prime}(\ov{t}) & = & \frac{-1}{2\ov{\cal J}\ov{B}_{\|}^{*}} \left( \pd{\ov{\Psi}^{*}}{\rho}\,\pd{\ov{B}}{\Lambda} - \pd{\ov{\Psi}^{*}}{\Lambda}\,\pd{\ov{B}}{\rho} \right), \label{eq:P_prime}
\end{eqnarray}
where the slow drift-precession time scale is clearly seen in Eq.~\eqref{eq:Phi_prime}, with $\nu_{\rm e}/\Omega_{\rm e} = \epsilon\,\ov{\mu}^{\frac{1}{2}}$ according to Eq.~\eqref{eq:nu_def}. For electrons (with $\mu B_{\rm e} = 100$ keV) or protons (with $\mu B_{\rm e} = 100$ MeV) at $r_{\rm e} = 2\,r_{\rm E}$, for example, we find $\nu_{\rm e} = 10.4$ Hz (electrons) or 7.67 Hz (protons) and $\nu_{\rm e}/\Omega_{\rm e} \simeq 1.5 \times 10^{-5}$ (electrons) or $2 \times 10^{-2}$ (protons). Hence, the azimuthal precession time scale is approximately $(\Omega_{\rm e}/\nu_{\rm e})$ longer than the bounce time scale (i.e., on the scale of a few hours compared to a few seconds). The bounce-center analysis \cite{Cary_Brizard_2009} of particle motion in an axisymmetric magnetic dipole field was carried out in Ref.~\cite{Duthoit_2010}. 

 \begin{figure}
\epsfysize=3.2in
\epsfbox{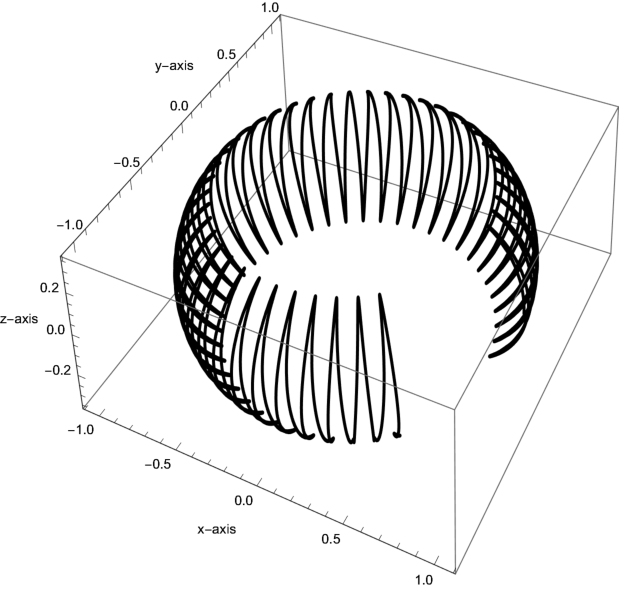}
\caption{Guiding-center orbit shown over several bounce periods based on the numerical integration of Eqs.~\eqref{eq:RLambda_prime}-\eqref{eq:P_prime} for $\nu_{\rm e}/\Omega_{\rm e} = 1/50$, with the initial conditions 
\eqref{eq:initial_gc}.}
\label{fig:bounce_precession}
\end{figure}

 \begin{figure}
\epsfysize=2in
\epsfbox{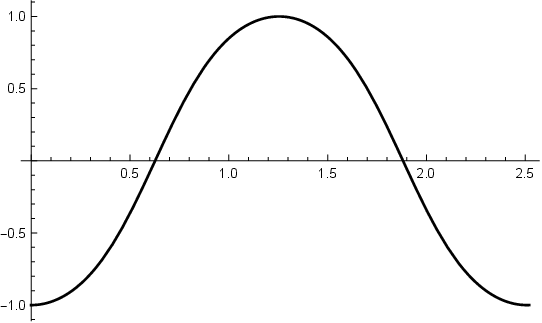}
\caption{Guiding-center parallel momentum $\ov{P}_{\|}(\ov{t})$ as a function of the normalized time $\ov{t} = \nu_{\rm e}t$ over one bounce period $T_{\rm b} \simeq 2.5$  for $\nu_{\rm e}/\Omega_{\rm e} = 1/50$, with the initial conditions \eqref{eq:initial_gc}.}
\label{fig:p_gc}
\end{figure}

\begin{figure}
\epsfysize=1.8in
\epsfbox{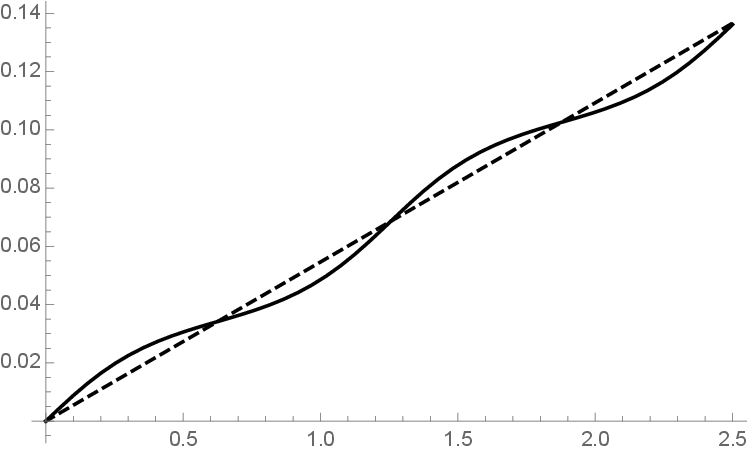}
\caption{Guiding-center azimuthal angle $\Phi(\ov{t})$ as a function of normalized time $\ov{t} = \nu_{\rm e}t$ (solid curve) over one bounce period for $\nu_{\rm e}/\Omega_{\rm e} = 1/50$, with the initial conditions \eqref{eq:initial_gc}. The dashed curve is a bounce-averaged straight line ($2\pi\ov{t}/T_{\rm d}$) whose slope is adjusted to a normalized drift-precession period $T_{\rm d} \simeq 115 \gg T_{\rm b} \simeq 2.5$.}
\label{fig:Phi_gc}
\end{figure}

Figures \ref{fig:bounce_precession}-\ref{fig:Phi_gc} show the numerical solutions of the guiding-center equations of motion \eqref{eq:RLambda_prime}-\eqref{eq:P_prime} with the initial conditions
\begin{equation}
(\rho_{0},\Phi_{0},\Lambda_{0},\ov{P}_{\|0}) \;=\; (1, 0, 0, -\,1),
\label{eq:initial_gc}
\end{equation}
and the normalized parameters $\ov{B}_{\rm t} = {\cal E} /\mu B_{\rm e} = 2$ (i.e., the equatorial pitch angle is $\alpha_{\rm e} = 3\pi/4$), and $\nu_{\rm e}/\Omega_{\rm e} = 2 \times 10^{-2}$ are used in Eq.~\eqref{eq:Phi_prime}. With these parameters \cite{Duthoit_2010}, the bounce latitude is $\Lambda_{\rm b} \simeq 0.4$, according to Eq.~\eqref{eq:lambda_b}, the normalized bounce period is $T_{\rm b} \simeq 2.5$ (see Fig.~\ref{fig:p_gc}), and the normalized precession period is $T_{\rm d} \simeq 115 \gg T_{\rm b}$ (see Fig.~\ref{fig:Phi_gc}). These guiding-center figures can be compared with the particle figures \ref{fig:orbit_par}, \ref{fig:v_par}, and \ref{fig:phi}, respectively, where the fast gyromotion time scale has been removed by the guiding-center transformation, while the bounce and drift-precession time scales remain intact.

\section{\label{sec:validity}Validity of the Guiding-center Approximation}

\begin{figure}
\epsfysize=2in
\epsfbox{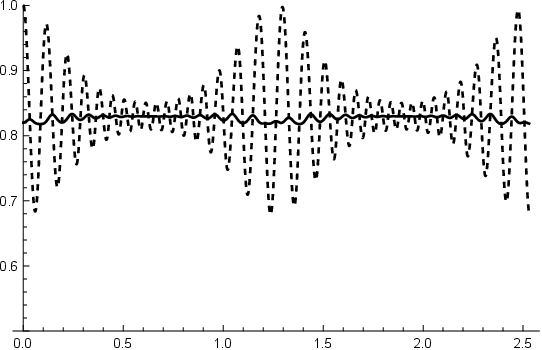}
\caption{Plots of the normalized magnetic moments $\ov{\mu}_{0}$ (dashed curve) and $\ov{\mu} = \ov{\mu}_{0} + \epsilon\,\ov{\mu}_{1}$ (solid curve) over a bounce period  $0 \leq \ov{t} = \epsilon\,\Omega_{\rm e}t \leq T_{\rm b} \simeq 2.5$ for $\epsilon = 1/50$, using the numerical solutions of the particle equations of motion \eqref{eq:rho_dot}-\eqref{eq:phi_dot} with initial conditions \eqref{eq:initial}.}
\label{fig:mu_01}
\end{figure}

We now proceed with a numerical comparison of the exact particle dynamics based on Eqs.~\eqref{eq:rho_dot}-\eqref{eq:phi_dot} with the guiding-center dynamics based on Eqs.~\eqref{eq:RLambda_prime}-\eqref{eq:P_prime}. In particular, we verify that the particle azimuthal canonical momentum \eqref{eq:P_phi} and the guiding-center azimuthal canonical momentum \eqref{eq:Pgc_phi} are indeed exact invariants of their respective dynamics. 

One of the important tests of the validity of the guiding-center approximation is concerned with how well the guiding-center magnetic moment $\mu$ is conserved by the particle dynamics. In Fig.~\ref{fig:mu_01} (dashed curve), we see that the normalized zeroth-order magnetic moment $\ov{\mu}_{0} = 2\,\mu_{0}B_{\rm e}/(\epsilon^{2}m\,r_{\rm e}^{2}\Omega_{\rm e}^{2})$ is not well conserved during a bounce period, with large-amplitude oscillations as the particle approaches the equatorial plane. When first-order corrections are added to the normalized magnetic moment $\ov{\mu} = \ov{\mu}_{0} + \epsilon\,\ov{\mu}_{1}$ (solid curve), however, only small oscillations are observed, even when the particle crosses the equatorial plane. The first-order correction $\mu_{1}$, which is calculated explicitly in App.~\ref{sec:gc_higher} for the magnetic dipole geometry, is therefore seen to play a crucial role in establishing the adiabatic invariance of the guiding-center magnetic moment.

\begin{figure}
\epsfysize=2in
\epsfbox{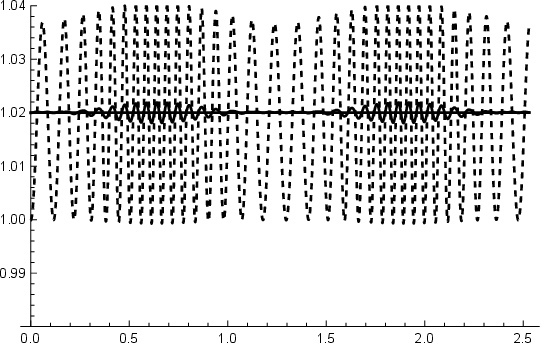}
\caption{Plots of the guiding-center pull-back of the normalized guiding-center azimuthal canonical angular momentum ${\sf T}_{\rm gc}\ov{P_{{\rm gc}\Phi}} = {\sf T}_{\rm gc}\ov{\Psi} - \epsilon^{2}\,(\ov{\mu}/2) \;{\sf T}_{\rm gc}b_{z}^{*}$ (solid curve) and the normalized particle azimuthal canonical angular momentum $\ov{P_{\varphi}}$ (black horizontal line), with $\ov{\Psi}(\ov{r},\lambda) = \cos^{2}\lambda/\ov{r}$ (dashed curve), over one bounce period $0 \leq \ov{t} = \epsilon\,\Omega_{\rm e}t \leq T_{\rm b} \simeq 2.5$ for $\epsilon = 1/50$, using the equations of motion \eqref{eq:rho_dot}-\eqref{eq:phi_dot} with the initial conditions \eqref{eq:initial}.}
\label{fig:Pphi_gc}
\end{figure}

Another important test explores the accuracy and faithfulness of the guiding-center azimuthal canonical momentum \eqref{eq:Pgc_phi} by transforming it back into particle phase space and use the particle dynamics to assess its invariance. This task allows us to determine the role of the higher-order curvature-corrected function $b_{z}^{*} = b_{z} + \frac{1}{2}\,K$ \cite{Tronko_Brizard_2015}, defined in Eq.~\eqref{eq:bz_*}, in contrast to the standard guiding-center term $b_{z}$ \cite{Littlejohn_1983}. We investigate how faithful the guiding-center azimuthal angular momentum $P_{{\rm gc}\Phi}$, defined by Eq.~\eqref{eq:Pgc_phi}, is to the exact azimuthal angular momentum $P_{\varphi}$, defined by Eq.~\eqref{eq:P_phi}. 

To perform a valid comparison, we need to transform the guiding-center azimuthal angular momentum $P_{{\rm gc}\Phi}$ back into particle phase space, with the help of the guiding-center pull-back operator ${\sf T}_{\rm gc}$. Here, using Eq.~\eqref{eq:X_gc}, the pull-back of an arbitrary guiding-center function $F_{\rm gc}({\bf X})$ is defined on particle phase space as 
\begin{equation}
{\sf T}_{\rm gc}F_{\rm gc}({\bf x},{\bf p}) \;=\; F_{\rm gc}\left({\bf x} - \vb{\rho}\right), 
\label{eq:FLR}
\end{equation}
which replaces the guiding-center position ${\bf X} = {\bf x} - \vb{\rho}$ with the particle position ${\bf x}$ and the finite-Larmor-radius (FLR) corrections $\vb{\rho} = \vb{\rho}_{0} + \vb{\rho}_{1} + \cdots$ involving the lowest-order gyroradius $\vb{\rho}_{0}
= \epsilon\,r_{\rm e}\ov{\vb{\rho}}_{0}$ and a correction $\vb{\rho}_{1} = \epsilon^{2} r_{\rm e}\ov{\vb{\rho}}_{1}$ associated with magnetic-field nonuniformity. Hence, the pull-back of the guiding-center azimuthal canonical angular momentum \eqref{eq:Pgc_phi} is
\begin{equation}
{\sf T}_{\rm gc}P_{{\rm gc}\Phi} \;=\; \frac{q}{c}{\sf T}_{\rm gc}\Psi \;-\;\left\{ \begin{array}{lr}
 \epsilon^{2}\,J \;{\sf T}_{\rm gc}b_{z}^{*} & ({\sf A}) \\
  & \\
 \epsilon^{2}\,J \;{\sf T}_{\rm gc}b_{z} & ({\sf B})
\end{array} \right. 
\label{eq:Tgc_Pphi}
\end{equation}
where $b_{z}^{*} = b_{z} + \frac{1}{2}\,K$ adds the guiding-center polarization (curvature) term $K$ to the standard guiding-center correction \cite{Littlejohn_1983,Belova_2003}. 

First, we see in Fig.~\ref{fig:Pphi_gc} that the guiding-center pull-back of the normalized guiding-center azimuthal canonical angular momentum (solid curve) 
\begin{equation} 
{\sf T}_{\rm gc}\ov{P_{{\rm gc}\Phi}} \;=\; {\sf T}_{\rm gc}\ov{\Psi} \;-\; \epsilon^{2}\,(\ov{\mu}/2) \;{\sf T}_{\rm gc}b_{z}^{*}
\label{eq:Tgc_Pphi_ov}
\end{equation}
is very close to the exact (normalized) particle azimuthal canonical angular momentum $\ov{P_{\varphi}}$ (solid horizontal line). The lowest-order component $\ov{\Psi}(\ov{r},\lambda) = \cos^{2}\lambda/\ov{r}$ is also shown as a dashed curve, which highlights the effects of the guiding-center pull-back operation ${\sf T}_{\rm gc}$ in Eq.~\eqref{eq:Tgc_Pphi_ov}.

\begin{figure}
\epsfysize=1.8in
\epsfbox{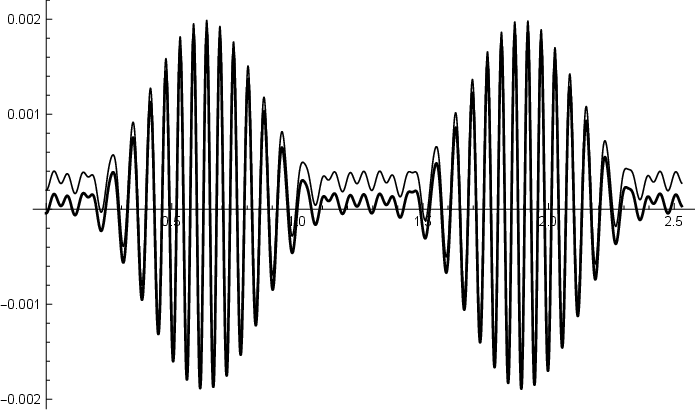}
\caption{Plot of the relative difference $({\sf T}_{\rm gc}P_{{\rm gc}\Phi})/P_{\varphi} - 1$ between the pull-back of the guiding-center azimuthal canonical angular momentum \eqref{eq:Tgc_Pphi} and the particle azimuthal canonical angular momentum \eqref{eq:P_phi} for $\epsilon = 1/50$, using the equations of motion \eqref{eq:rho_dot}-\eqref{eq:phi_dot} with the initial conditions \eqref{eq:initial}. The dark curve uses the guiding-center polarization correction (A) in Eq.~\eqref{eq:Tgc_Pphi} while the light curve only shows the standard correction (B).}
\label{fig:psi_gc}
\end{figure}

Second, the relative difference $({\sf T}_{\rm gc}P_{{\rm gc}\Phi})/P_{\varphi} - 1$ between the pull-back of the guiding-center azimuthal canonical angular momentum \eqref{eq:Tgc_Pphi} and the particle azimuthal canonical angular momentum \eqref{eq:P_phi} is shown in Fig.~\ref{fig:psi_gc}, where excellent agreement is observed when the guiding-center polarization curvature term $\frac{1}{2}\,K$ (dark curve) is added to the standard correction $b_{z}$ (light curve). Hence, the guiding-center polarization correction introduced in Ref.~\cite{Tronko_Brizard_2015} yields a guiding-center azimuthal canonical angular momentum that is faithful to the particle azimuthal canonical angular momentum. 

\begin{figure}
\epsfysize=1.8in
\epsfbox{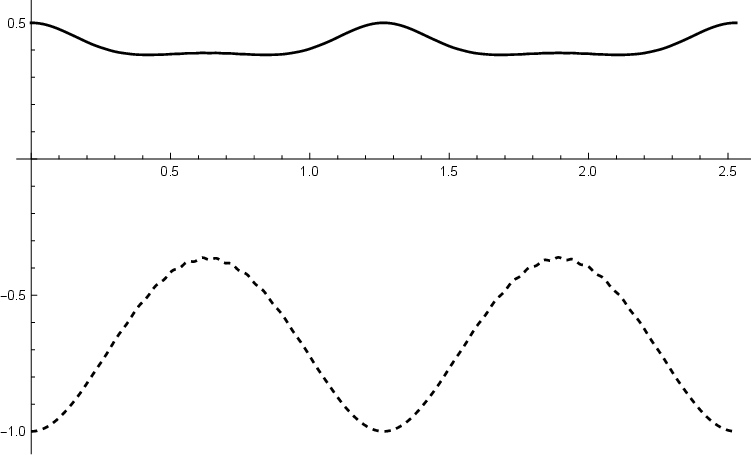}
\caption{Plots of ${\sf T}_{\rm gc}b_{z}^{*}$ (solid curve) and ${\sf T}_{\rm gc}b_{z}$ (dashed curve) over a (normalized) bounce period for $\epsilon = 1/50$, using the equations of motion \eqref{eq:rho_dot}-\eqref{eq:phi_dot} with the initial conditions \eqref{eq:initial}.} 
\label{fig:mu_bz_gc}
\end{figure}

The reason for the success of this polarization correction is seen in Fig.~\ref{fig:mu_bz_gc}, where the positive guiding-center pull-back ${\sf T}_{\rm gc}b_{z}^{*}$ brings a correction toward $\ov{P}_{\varphi}$, according to Eq.~\eqref{eq:Tgc_Pphi_ov}, while the negative guiding-center pull-back ${\sf T}_{\rm gc}b_{z}$ brings a correction away from $\ov{P}_{\varphi}$. Hence, based on Fig.~\ref{fig:b_z}, it is expected that guiding-center polarization effects will play a greater role for deeply-trapped charged particles $(\lambda_{\rm b} \ll 1)$, for which $b_{z}^{*} \simeq 1/2$ while $b_{z} \simeq -1$, in establishing a faithful guiding-center representation for regular particle orbits in magnetic dipole geometry. Using the bounce latitude  $\lambda_{\rm b} \simeq 0.4$ (see the center frame of Fig.~\ref{fig:rho_lambda_psi}), for example, the faithful role of the guiding-center polarization is noticeable in Figs.~\ref{fig:psi_gc}-\ref{fig:mu_bz_gc}. For $\lambda_{\rm b} > 1$, on the other hand, we expect the difference between the guiding-center pull-backs ${\sf T}_{\rm gc}b_{z}^{*} > {\sf T}_{\rm gc}b_{z}$ to be small and, thus, the role of the guiding-center polarization in providing a faithful guiding-center representation of charged-particle dynamics in a magnetic dipole field is expected to be less important.

\section{Summary}

In previous work \cite{Brizard_2017}, we showed that the guiding-center approximation was valid in a straight magnetic field with constant perpendicular magnetic gradient, even in the presence of strong gradients. In addition, based on the existence of an exact analytical solution for the particle orbits, this work also confirmed that the guiding-center polarization corresponded exactly with an orbit-averaged particle displacement.

In the present work, we extended our investigation of the validity of the guiding-center approximation in describing charged single-particle motion in a nonuniform magnetic field. Here, we considered regular particle orbits in azimuthally symmetric magnetic dipole geometry, in which the azimuthal angular canonical momentum is conserved and the guiding-center magnetic moment is an adiabatic invariant. We successfully validated the guiding-center approximation in describing particle motion in an azimuthally symmetric magnetic dipole field provided higher-order guiding-center corrections are taken into account, which had already been noted for the case of an axisymmetric tokamak magnetic field \cite{Belova_2003}. In particular, the guiding-center polarization correction in the guiding-center azimuthal angular canonical momentum, not taken into account in the standard guiding-center approximation \cite{Littlejohn_1983}, proved crucial in establishing a faithful guiding-center representation for regular particle orbits in magnetic dipole geometry.

\vspace*{0.1in}

\begin{center}
{\bf Data Availability Statement}
\end{center}

The Mathematica code used to generate the plots in the present manuscript is available upon request.

\appendix

 \section{\label{sec:gc_higher}Higher-order Guiding-center Approximation}
 
 In this Appendix, we will calculate explicit guiding-center corrections that arise at first-order in the magnetic-field nonuniformity associated with magnetic-field gradients $(\nabla\ln B)$ and magnetic-field curvature $(\vb{\kappa} = \bhat\bdot\nabla\bhat)$. 
 
First, we explore the adiabatic invariance of the lowest-order magnetic moment 
\begin{equation}
\mu_{0} = \frac{m|{\bf v}_{\bot}|^{2}}{2B} = \left(\epsilon^{2}m r_{\rm e}^{2}\Omega_{\rm e}^{2}/2 B_{\rm e}\right)\;\frac{|\ov{\bf v}_{\bot}|^{2}}{\ov{B}}. 
\end{equation}
First, we find that $\dot{\mu}_{0} = -\,\mu_{0}\,{\bf v}\bdot\nabla\ln B + m\dot{\bf v}_{\bot}\bdot
{\bf v}_{\bot}/B$, which can ultimately be expressed as
\begin{eqnarray} 
\dot{\mu}_{0} & = & -\,\mu_{0}\,{\bf v}\bdot\nabla\ln B + \frac{m}{B}\left(\dot{\bf v} - \dot{v}_{\|}\,\bhat - v_{\|}\,{\bf v}\bdot\nabla\bhat\right)\bdot{\bf v}_{\bot} \nonumber \\
 &=& -\,\mu_{0}\,{\bf v}\bdot\nabla\ln B - \frac{mv_{\|}}{B}\;{\bf v}\bdot\nabla\bhat\bdot{\bf v}_{\bot} \nonumber \\
 & = &  -\,\frac{v_{\|}}{B}\left( \mu_{0}\,\bhat\bdot\nabla B \;+\frac{}{} m\,{\bf v}_{\bot}\bdot\nabla\bhat\bdot{\bf v}_{\bot} \right) \nonumber \\
  &&-\; \frac{{\bf v}_{\bot}}{B} \bdot\left( \mu_{0}\,\nabla B \;+\; m\,v_{\|}^{2}\,\bhat\bdot\nabla\bhat\right).
 \label{eq:mu0_dot}
 \end{eqnarray}
Although $\dot{\mu}_{0} \neq 0$ for general magnetic fields, its average over the fast gyromotion time scale yields
 \[ \langle\dot{\mu}_{0}\rangle \;=\; -\,\mu_{0}\,v_{\|} \left( \bhat\bdot\nabla \ln B \;+\frac{}{} \nabla\bdot\bhat\right) \;=\; -\,\frac{\mu_{0}\,v_{\|}}{B}\;(\nabla\bdot{\bf B}), \]
 where $\langle{\bf v}_{\bot}\rangle = 0$ and $\langle{\bf v}_{\bot}\bdot\nabla\bhat\bdot{\bf v}_{\bot}\rangle = (\nabla\bdot\bhat)\;|{\bf v}_{\bot}|^{2}/2$. Since magnetic fields are divergenceless, we immediately find that 
 $\langle\dot{\mu}_{0}\rangle = 0$, i.e., $\mu_{0}$ is an invariant over time scales that are long compared to the fast gyromotion time scale.

  \subsection{First-order correction to $\mu_{0}$}
  
The normalized guiding-center magnetic moment can be constructed $\ov{\mu} = \ov{\mu}_{0} + \epsilon\,\ov{\mu}_{1} + \cdots$ as an asymptotic expansion in powers of $\epsilon$, where $\ov{\mu}_{1}$ is a correction that involves the nonuniformity of the magnetic field. For a general magnetic field, the first-order correction to the magnetic moment is \cite{Tronko_Brizard_2015}
 \begin{eqnarray}
 \mu_{1} &=& \left(\mu_{0}\nabla_{\bot}\ln B + \frac{p_{\|}^{2}\,\vb{\kappa}}{m\,B}\right)\vb{\cdot}\vb{\rho}_{0} - \mu_{0}\varrho_{\|} \left(\tau \;+\frac{}{} \alpha_{1}\right) \nonumber \\
  &\equiv& (\epsilon^{2}m r_{\rm e}^{2}\Omega_{\rm e}^{2}/2 B_{\rm e})\,\epsilon\,\ov{\mu}_{1},
 \label{eq:mu1_def}
 \end{eqnarray}
 where $\alpha_{1} = \frac{1}{2}\,\tau - \wh{\bot}\bdot\nabla\bhat\bdot\wh{\rho}$. For magnetic dipole geometry, where $\tau = \bhat\bdot\nabla\btimes\bhat = 0$ and $\nabla_{\bot}\ln B = \vb{\kappa}$, we find
 \[ -\,\mu_{0}\varrho_{\|}\alpha_{1} \;=\; \frac{p_{\|}}{2B}\;\left(\frac{d\bhat}{dt} \;-\; v_{\|}\;\vb{\kappa}\right)\bdot\vb{\rho}_{0},  \]
 so that Eq.~\eqref{eq:mu1_def} becomes
  \begin{equation}
 \mu_{1} \;=\; \left(\mu_{0} \;+\; \frac{p_{\|}^{2}}{2m\,B}\right)\vb{\kappa}\bdot\vb{\rho}_{0} \;+\; \frac{p_{\|}}{2B}\;\frac{d\bhat}{dt}\bdot\vb{\rho}_{0},
 \label{eq:mu1}
 \end{equation}
 where $\vb{\kappa}\bdot\vb{\rho}_{0} = -\;\epsilon\,\varphi^{\prime}K/\ov{B}$. Hence, the normalized magnetic-moment correction $\ov{\mu}_{1}$ in magnetic dipole geometry is expressed as
 \begin{equation} 
 \ov{\mu}_{1} \;=\; -\,\left(\ov{\mu}_{0} \;+\; \frac{\ov{p}_{\|}^{2}}{\ov{B}}\right) \frac{\varphi^{\prime}}{\ov{B}}\;K \;+\; \frac{\ov{p}_{\|}}{\ov{B}}\;\bhat^{\prime}\bdot\ov{\vb{\rho}}_{0}.
 \label{eq:ov_mu_1}
 \end{equation}
 Figure \ref{fig:mu_01} shows how important the first-order correction \eqref{eq:mu1} is in establishing the adiabatic invariance of magnetic moment. Indeed, using $\dot{\mu}_{1} = \Omega\,\partial\mu_{1}/\partial\zeta$ to lowest order and $\dot{\vb{\rho}}_{0} = \Omega\,\partial\vb{\rho}_{0}/\partial\zeta = {\bf v}_{\bot}$, we see that
 \begin{widetext}
 \begin{eqnarray}
 \dot{\mu}_{0} + \dot{\mu}_{1} &=& -\,\frac{v_{\|}}{B}\left( \mu_{0}\,\bhat\bdot\nabla B \;+\frac{}{} m\,{\bf v}_{\bot}\bdot\nabla\bhat\bdot{\bf v}_{\bot} \right) \;-\; \frac{{\bf v}_{\bot}}{B} \bdot\left( \mu_{0}\,\nabla_{\bot} B \;+\; m\,v_{\|}^{2}\,\bhat\bdot\nabla\bhat\right) 
 \nonumber \\
   &&+\; \left(\mu_{0}\;\nabla_{\bot}\ln B \;+\; \frac{p_{\|}^{2}\,\vb{\kappa}}{m\,B}\right)\bdot{\bf v}_{\bot} \;+\; \frac{p_{\|}}{2B}\;\left({\bf v}_{\bot}\bdot\nabla\bhat\bdot{\bf v}_{\bot} - \pd{{\bf v}_{\bot}}{\zeta}\bdot\nabla\bhat\bdot\pd{{\bf v}_{\bot}}{\zeta}\right) 
   \nonumber \\
    &=& -\,\frac{v_{\|}}{B}\;\mu_{0}\,\bhat\bdot\nabla B \;-\; \frac{p_{\|}}{2B}\;\left({\bf v}_{\bot}\bdot\nabla\bhat\bdot{\bf v}_{\bot} + \pd{{\bf v}_{\bot}}{\zeta}\bdot\nabla\bhat\bdot\pd{{\bf v}_{\bot}}{\zeta}\right) \nonumber \\
     &=& -\,\frac{\mu_{0}\,v_{\|}}{B} \left( \bhat\bdot\nabla B \;+\frac{}{} B\;\nabla\bdot\bhat\right) \;=\;  -\,\frac{\mu_{0}\,v_{\|}}{B}\;(\nabla\bdot{\bf B}) \;=\; 0,
 \end{eqnarray}
\end{widetext} 
and, thus, the guiding-center magnetic moment $\mu$ is conserved up to first-order effects in magnetic-field nonunifornity, without the need of gyroangle-averaging.

\begin{figure}
\epsfysize=2in
\epsfbox{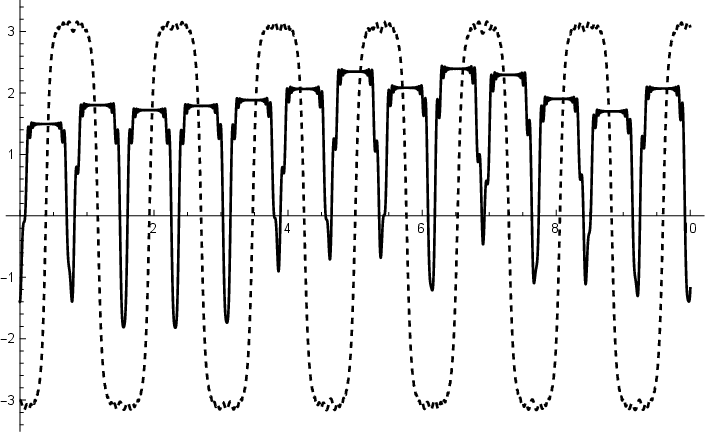}
\caption{Plots of the normalized magnetic moment $\ov{\mu} = \ov{\mu}_{0} + \epsilon\,\ov{\mu}_{1}$ (solid curve) and the normalized parallel particle velocity $\ov{v}_{\|}$ (dashed curve) over several bounce periods for $\epsilon = 1/50$, using the equations of motion \eqref{eq:rho_dot}-\eqref{eq:phi_dot} and the modified initial conditions \eqref{eq:initial} with $\lambda_{0}^{\prime} = 3$. The magnetic moment is a good invariant when the particle is near turning points (where the magnetic field is the strongest), while the invariant level fluctuates after each equatorial crossing.}
\label{fig:mu_01_chaos}
\end{figure}

Lastly, we note that not all charged-particle orbits in a nonuniform magnetic field lead to the adiabatic invariance of the guiding-center magnetic moment. Indeed, the breakdown of the adiabatic invariance of the guiding-center magnetic moment is well documented on theoretical grounds \cite{Dragt_Finn_1976,Bernstein_Rowlands_1976,Cohen_Rowlands_1978,Neishtadt_2019} as well as being numerically investigated for magnetic dipole geometry \cite{Dragt_Finn_1976,Murakami_1990,Kuznetsov_Yushkov_2002,Xie_Liu_2020} and axisymmetric tokamak geometry \cite{Carlsson_2001,Koleskichenko_2002,Yavorskij_2002,Escande_Sattin_2021}. In our numerical analysis of charged-particle orbits in magnetic dipole geometry, for example, Fig.~\ref{fig:mu_01_chaos} shows what happens when the initial conditions \eqref{eq:initial} are modified with $\lambda_{0}^{\prime} = 3$, which increases the normalized equatorial velocity $|\ov{v}_{\|{\rm e}}| \simeq 3$. Here, we see that the magnetic moment is a good invariant when the particle is near turning points (where the magnetic field is the strongest), while the invariant level fluctuates after each equatorial crossing. While this charged particle is still magnetically confined, a further increase in $\lambda_{0}^{\prime}$ leads to loss of magnetic confinement as a result of the destruction of the adiabatic invariance of the magnetic moment. Additional details about the breakdown of adiabatic invariance in axisymmetric magnetic dipole geometry can be found in Refs.~\cite{Murakami_1990} and \cite{Xie_Liu_2020}.

\subsection{First-order correction to $\vb{\rho}_{0}$}

When FLR corrections \eqref{eq:X_gc} need to be calculated up to first order in magnetic-field nonuniformity, we must evaluate the pull-back of the guiding-center position up to second order in $\epsilon$ (now appearing as an ordering parameter associated with the rescaling $q \rightarrow q/\epsilon$):
\begin{eqnarray}
{\sf T}_{\rm gc}{\bf X} &=& {\bf x} + \epsilon\,G_{1}^{\bf x} + \epsilon^{2} \left(G_{2}^{\bf x} \;+\; \frac{1}{2}\,{\sf G}_{1}\cdot\exd G_{1}^{\bf x}\right) + \cdots \nonumber \\
 &\equiv&  {\bf x} - \vb{\rho} \;=\; {\bf x} - \epsilon\,\vb{\rho}_{0} \;-\; \epsilon^{2}\,\vb{\rho}_{1} + \cdots
 \end{eqnarray}
 where, using the components $(G_{2}^{\bf x}, G_{1}^{\mu}, G_{1}^{\zeta})$ calculated in Ref.~\cite{Tronko_Brizard_2015}, we obtain the first-order gyroradius
 \begin{eqnarray}
 \vb{\rho}_{1} &=& -\,G_{2}^{\bf x} \;-\; \frac{1}{2}\,\vb{\rho}_{0}\bdot\nabla\vb{\rho}_{0} \;+\; \frac{1}{2}\,\left(G_{1}^{\mu}\,\pd{\vb{\rho}_{0}}{\mu} + G_{1}^{\zeta}\,\pd{\vb{\rho}_{0}}{\zeta}\right) \nonumber \\
  &\equiv& \rho_{1\|}\,\bhat \;+\; \frac{1}{2}\,\left(\vb{\rho}_{0}\bdot\nabla\ln B\right) \vb{\rho}_{0} \;-\; \frac{\mu B}{2m \Omega^{2}}\;\vb{\kappa}.
 \end{eqnarray} 
Here, $\vb{\rho}_{0}\bdot\nabla\ln B = -\,\dot{\varphi}\,K/\Omega$ and the parallel component is defined as
 \[ \rho_{1\|} = \frac{5}{4} \left[ \frac{\mu B}{2m \Omega^{2}}\left(\nabla\bdot\bhat\right) - \frac{v_{\|}}{\Omega}\,K\,\vb{\rho}_{0}\bdot\nabla\varphi \right] - \frac{3}{4\Omega^{2}}\;\frac{d\bhat}{dt}\bdot{\bf v}. \]
This first-order gyroradius correction can then be used in the guiding-center pull-back \eqref{eq:FLR}.
 
We now calculate the guiding-center gyroradius $\vb{\rho}_{\rm gc} \equiv {\sf T}_{\rm gc}^{-1}{\bf x} - {\bf X} = {\sf T}_{\rm gc}^{-1}\vb{\rho}$, which is defined in terms of the guiding-center push-forward of the particle gyroradius into guiding-center phase space, where $\vb{\rho}_{{\rm gc}0} = \vb{\rho}_{0}$ and $\vb{\rho}_{{\rm gc}1} = \vb{\rho}_{1} - {\sf G}_{1}\cdot\exd\vb{\rho}_{0}$. In Ref.~\cite{Tronko_Brizard_2015}, it was shown that the guiding-center polarization can be constructed explicitly from the guiding-center transformation, which is defined as the gyroangle-averaged expression \cite{Brizard_2013}
 \begin{eqnarray}
 \vb{\pi}_{\rm gc} &\equiv& q\,\langle\vb{\rho}_{{\rm gc}1}\rangle \;-\; \nabla\bdot\left\langle\frac{q}{2}\,\vb{\rho}_{0}\vb{\rho}_{0}\right\rangle  
 \label{eq:gc_pol} \\
  &=& -\;\frac{q}{m\Omega^{2}} \left( \mu\,B\;\nabla_{\bot}\ln B \;+\; \frac{p_{\|}^{2}}{m}\,\vb{\kappa}\right) \;=\; \frac{q\bhat}{\Omega}\btimes\frac{d{\bf X}}{dt}, \nonumber
\end{eqnarray}
thereby recovering the standard expression derived by Kaufman \cite{Kaufman_1986}. The higher-order guiding-center corrections \cite{Tronko_Brizard_2015} leading to Eq.~\eqref{eq:gc_pol} also yield the correction $b_{z}^{*}$ in Eq.~\eqref{eq:Tgc_Pphi}, which ensures a faithful guiding-center representation for regular charged-particle orbits in an axisymmetric magnetic dipole field.

\end{document}